\def \SAIT #1 #2 {{\em Mem.\ Soc.\ Astron.\ It.\/} {\bf #1}, #2}
\def \MESS #1 #2 {{\em The Messenger\/} {\bf #1}, #2}
\def \ASTRNACH #1 #2 {{\em Astron. Nach.\/} {\bf #1}, #2}
\def \AAP #1 #2 {{\em Astron. Astrophys.\/} {\bf #1}, #2}
\def \AAL #1 #2 {{\em Astron. Astrophys. Lett.\/} {\bf #1}, L#2}
\def \AAR #1 #2 {{\em Astron. Astrophys. Rev.\/} {\bf #1}, #2}
\def \AAS #1 #2 {{\em Astron. Astrophys. Suppl. Ser.\/} {\bf #1}, #2}
\def \AJ #1 #2 {{\em Astron. J.\/} {\bf #1}, #2}
\def \ANNREV #1 #2 {{\em Ann. Rev. Astron. Astrophys.\/} {\bf #1}, #2}
\def \APJ #1 #2 {{\em Astrophys. J.\/} {\bf #1}, #2}
\def \APJL #1 #2 {{\em Astrophys. J. Lett.\/} {\bf #1}, L#2}
\def \APJS #1 #2 {{\em Astrophys. J. Suppl.\/} {\bf #1}, #2}
\def \APSS #1 #2 {{\em Astrophys. Space Sci.\/} {\bf #1}, #2}
\def \ASR #1 #2 {{\em Adv. Space Res.\/} {\bf #1}, #2}
\def \BAIC #1 #2 {{\em Bull. Astron. Inst. Czechosl.\/} {\bf #1}, #2}
\def \JSQRT #1 #2 {{\em J. Quant. Spectrosc. Radiat. Transfer\/} {\bf #1}, #2}
\def \MN #1 #2 {{\em Mon. Not. R. Astr. Soc.\/} {\bf #1}, #2}
\def \MEM #1 #2 {{\em Mem. R. Astr. Soc.\/} {\bf #1}, #2}
\def \PLR #1 #2 {{\em Phys. Lett. Rev.\/} {\bf #1}, #2}
\def \PASJ #1 #2 {{\em Publ. Astron. Soc. Japan\/} {\bf #1}, #2}
\def \PASP #1 #2 {{\em Publ. Astr. Soc. Pacific\/} {\bf #1}, #2}
\def \NAT #1 #2 {{\em Nature\/} {\bf #1}, #2}

\documentstyle{memsait}
\input epsf.sty
\begin{opening}
\title{The VIRMOS mask manufacturing tools: a) Mask preparation software}
\author{B.Garilli$^1$, D.Bottini$^1$, L.Tresse$^2$, O.Le Fevre$^3$, M.Saisse$^3$, G.Vettolani$^2$}
\institute{$^1$Istituto di Fisica Cosmica "G. Occhialini", CNR, Milano, \\
$^2$Istituto di radioastronomia, CNR, Bologna, \\
$^3$Laboratoire d'Astronomie Spatiale, CNRS, Marseille}
\date{} % DO NOT INSERT ANY DATE HERE !!!
\end{opening}

\begin{document}

%\oddpagefooter{\sf Mem. S.A.It., Vol. ??, ??}{}{\thepage}
%\evenpagefooter{\thepage}{}{\sf Mem. S.A.It., Vol. ??, ??}
\oddpagefooter{}{}{} % LEAVE AS IT IS !
\evenpagefooter{}{}{} % LEAVE AS IT IS !
\ 
\bigskip

\begin{abstract}
VIMOS and NIRMOS are two multi Object Spectrographs being developed for the VLT. Their main scientific characteristic is the high multiplexing
capability, allowing to obtain up to 800 spectra per exposure. To achieve such a high efficiency, a number of dedicated tools are necessary, from software to hardware. In part a), we describe the software tools
which will be provided to successful proponents to prepare observations. Paper b) will deal with the technical aspects of mask manufacturing
\end{abstract}

\section{Introduction}
The Visual and Near Infrared Multi Object Spectrographs (VIMOS and
NIRMOS respectively) are two spectrographs being developed by the Franco-Italian Consortium VIRMOS for the VLT (LeFevre et al, 1998). They are due in Operation for Spring 2000 and Spring 2001 respectively. Both will have imaging capabilities and an Integral Field Unit, allowing for full field spectroscopy over 1 arcmin$^2$. 
VIMOS field of view will be composed of four quadrants of 7x8 arcmin,
for a total of 224 arcmin$^2$. Spectroscopy will be done using user defined slits
cut into INVAR masks, one mask per quadrant.
Multiobject spectroscopy will be possible in
either High resolution mode ($R\sim2500$) and low resolution mode ($R\sim200$).
In the low resolution mode, the design of the instrument has been done in such a way as to exploit spectra stacking, so that the maximum number
of spectra per exposure can be obtained. Given the CCD size (2048x4096
pixels) and the grism resolution, up to 200 spectra per quadrant can be stacked in one image (a total of 800 spectra per exposure). 

To exploit these capabilities, tools must be provided to
the astronomer to automatize the object selection procedure,
and to produce masks for the spectroscopic observations. 
In the case of the VLT, ESO has provided a common
scenario (the so called "VLT Data Flow System")
within which we had to accomodate the particular needs of 
VIMOS. In this contribution we will outline the software we are going
to provide for handling VIMOS (and NIRMOS)
spectroscopic observations. In part b), the choice of
the mask manufacturing unit and its handling are outlined

\section{Operations Overview}
VIMOS observations will be typically performed in two phases:
an observation in imaging mode of the field, in order to get the field object positions, and the spectroscopic observation of an appropriate number of the previously selected objects. After the imaging observation, but before the spectroscopic one, 4 masks, one for each quadrant, shall be manufactured.
During the manufacturing process, the slits corresponding to the objects to be
spectroscopically observed will be cut.

Before the spectroscopic observations start, all masks to be used during the night will be loaded into the instrument by putting them
into an appropriate container (Instrument Cabinet). The
Instrument Cabinets will be brought back and forth from the instrument
to the mask manufacturing building, where the Laser Cutting Machine will
be installed, together with its own
control devices.

The Mask Preparation Software shall perform the selection of the objects to be spectroscopically observed, the slit positioning,
the transformation from astronomical to laser machine
coordinates and will interact with MMU control software.

MPS is divided into two parts: MPS-P2PP (Phase 2 Proposal preparation), as
it is tied with P2PP, and MPS-IWS (Instrument WorkStation), as it
shall run exclusively on the IWS in Paranal.

MPS-P2PP provides the astronomer with tools for the selection of the objects to be spectroscopically observed and an algorithm for
the slit positioning in such a way as to get the most effective solution
in terms of number of spectra per field. Slit dimensions and positions will
be stored in Aperture Definition Files. ADFs will be included in the appropriate Observation Block during the spectroscopic P2PP phase.
When ADFs get in Paranal, ESO Observation Handling System will
extract the ADFs from Observations Blocks and pass them to MPS-IWS for mask manufacturing

MPS-IWS is responsible for converting slit coordinates from
astronomical (RA \& Dec) to manufacturing machine coordinates, assigning
a code identification
for each mask to be manufactured, sending such files to the Mask Manufacturing Control Unit (Mask Manufacturing Order), and receiving
acknowledge when
the masks have been completed and stored (Mask Manufacturing Report). 
Such information will be passed
back to OHS, and from that moment the observation can be scheduled 
for the following nights.

When a spectroscopic observation is scheduled, OHS sends to MPS-IWS a Mask Insertion Order, i.e. the list of masks to be put into the
Instrument Cabinet for subsequent nights. 
Such order is reformatted and passed on
to MMCU. The acknowledge from MMCU is passed backed to OHS under the form of a Mask Insertion Report. Only observations for which the 
corresponding masks
exist and are in the instrument cabinet can be performed

At the moment of observation, mask alignment onto the focal plane
is performed by a procedure implemented at Observation template level.
The mask is put in the focal plane, then a short exposure
without dispersion element is performed. The 4 masks have some
holes, corresponding to "reference objects", of pre-defined dimension
(at moment 5x5 arcsec are foreseen). Given the VLT pointing
accuracy, in the resulting image the reference objects should fall within the
holes, though not necessarily centered. An iterative procedure takes care of 
computing the pointing shift needed to center reference objects in
reference holes. Such procedure can be either manual, or even fully automatic.
When the reference objects are centered in their respective holes, within a
predefined accuracy, the dispersion element is introduced and the
observation can start.

As a last step, when an observation has been successfully carried out
(the quality Control ESO Pipeline is in charge of assuring that), OHS sends to MPS-IWS a Mask Discarding Order, which is re-sent to MMCU.
Once a mask has been discarded, also the OB can be deleted.

In figure 1, the Data flow in the case of Manufacturing is schematized.

\begin{figure}
\epsfxsize=9cm % fix the x-dimension and scales y-dim. to x-dim.
% Feel free to do the choice you prefer but do not exceed the x-dimension
% of the text lines
\hspace{1.5cm}\epsfbox{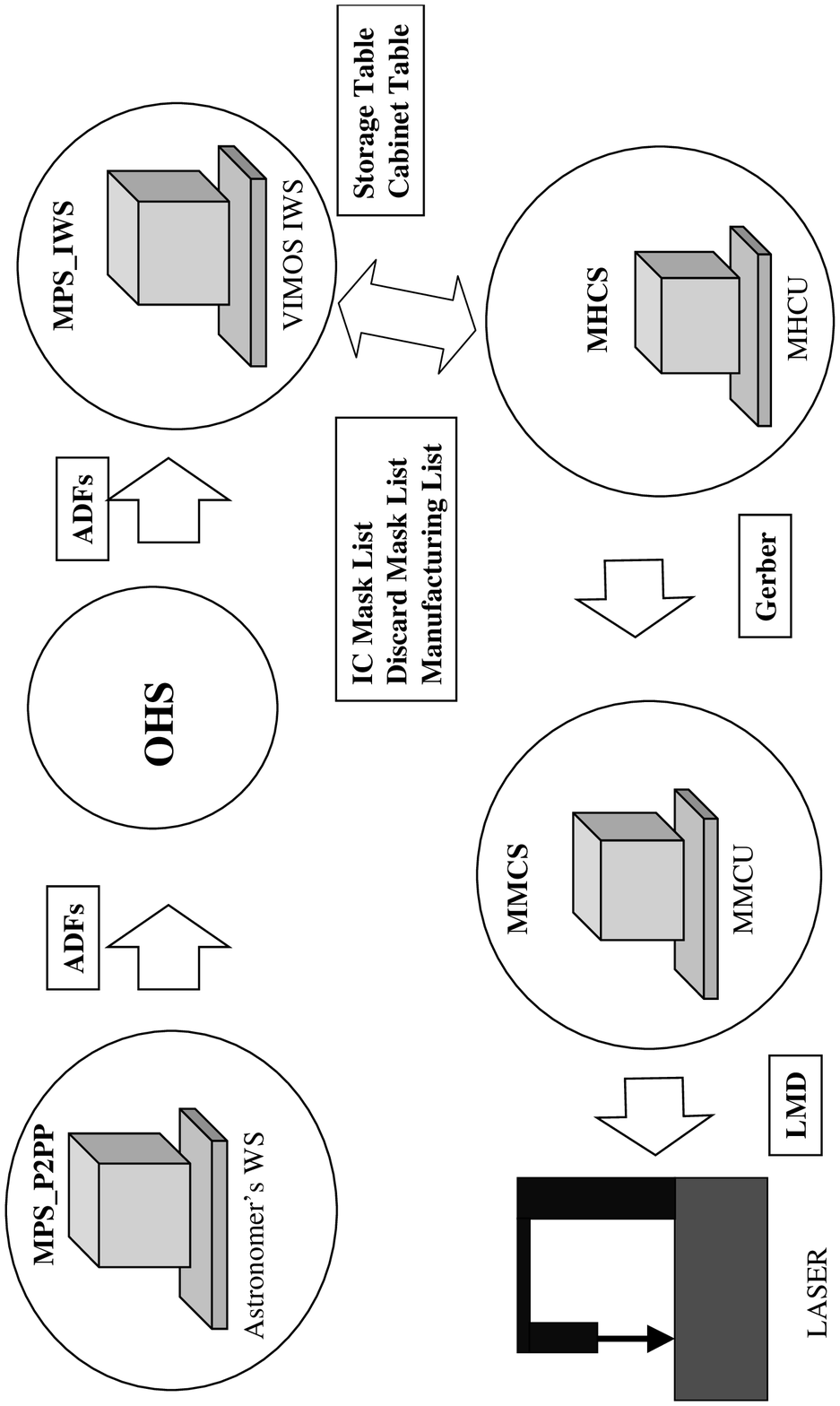} 
%for centering: act on hspace argument 
\caption[h]{Mask Preparation Data Flow}
\end{figure}

Here, we will not further deal with MPS-IWS, as it basically
acts as a link bewteen ESO OHS and the PC-based machine control
computers.

\section{User's requirements and observational constraints}
In designing MPS-P2PP, we have taken into considerations the following
user's requirements

$-$ To exploit multiplexing capabilities, an automatized way for choosing objects must be provided

$-$ Manual choice of some particularly interesting objects must be possible

$-$ Manual exclusion of particular objects must be possible

$-$ Manual definition of arbitrary shaped slits (i.e. curved slits) must be implemented

$-$ Interactions must rely on a user friendly graphical user interface

$-$ Choice of objects starting from a user provided catalogue 
(including at least Right Ascension and Declination for each object)
 must be possible

On top of these, the observation procedure outlined above, also requires 

$-$ Manual choice of reference objects to be used for mask alignment onto the focal plane

The need of a graphical interaction has led us to adopt one of the
available display systems. Our first choice has been MIDAS, which
gives the advantage of supporting table handling. A working
version of MPS-P2PP MIDAS based is already available.
To keep similarity with FORS FIMS package, though, we are now
implementing a second version of MPS-P2PP, based on Skycat for the image display and catalogue overlay features

The algorithm for automatic object choice has been totally developed
in house.

\section{Slit Positioning Optimization Code: SPOC}
SPOC is a C program which, taking into account the initial list
of user's preferred objects, the intrument constraints and
the constraints given by the wish of having well
reduced data,
finds the best solution in terms of number of slits. 
Instrument constraints are mainly given by the spectrum length, coupled with CCD size. Slits must be positioned so that the dispersed spectrum falls
entirely within the detector. 
To ease data reduction, we have to allow for a certain 
amount of sky on each side of the object falling in the slit, we
must avoid spectra superposition and we must take into account
spectra higher order superpositions.
In fig. 2 we illustrate different configurations

\begin{figure}
\epsfxsize=9cm % fix the x-dimension and scales y-dim. to x-dim.
% Feel free to do the choice you prefer but do not exceed the x-dimension
% of the text lines
\hspace{1.5cm}\epsfbox{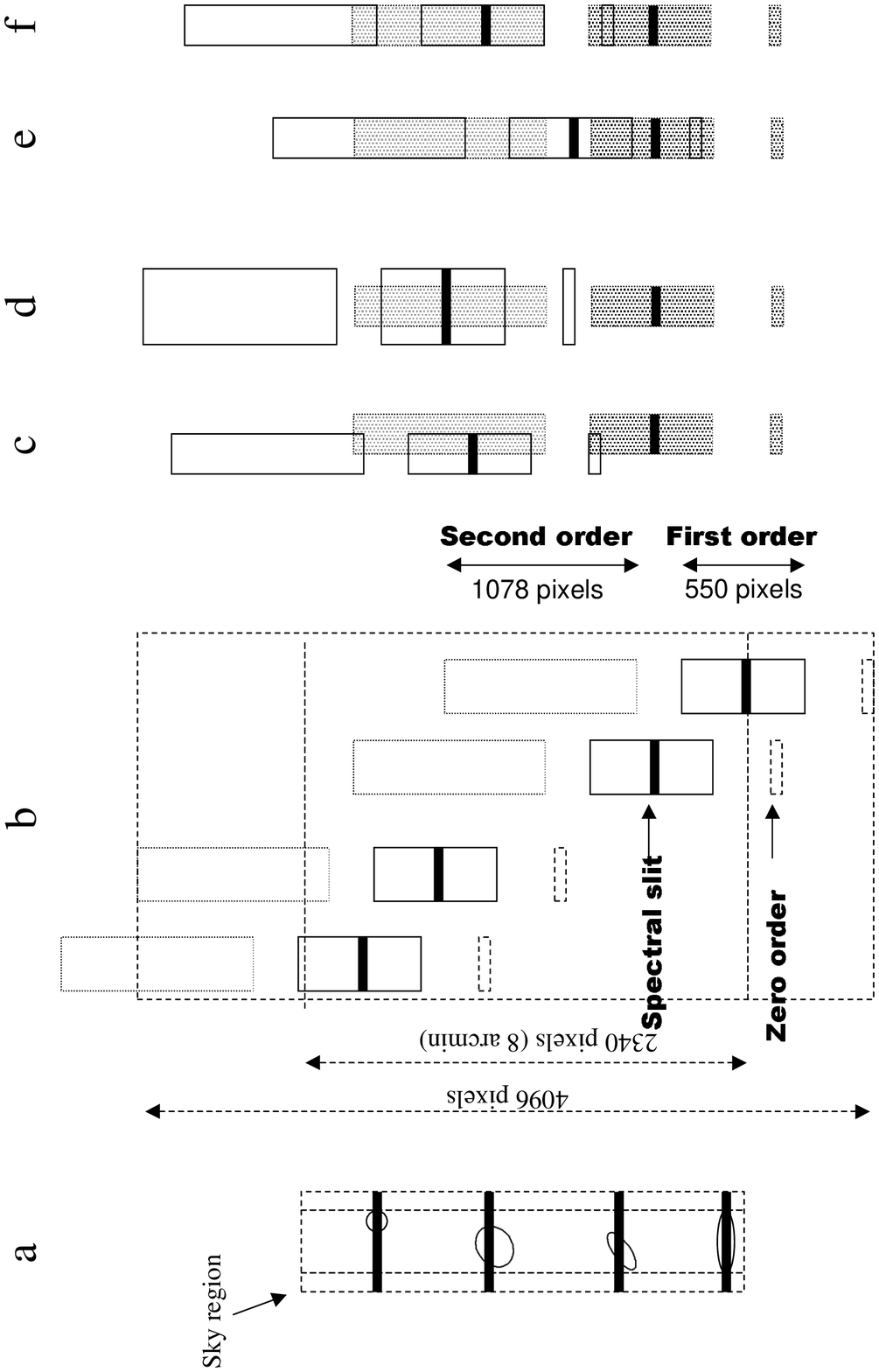} 
%for centering: act on hspace argument 
\caption[h]{Slit positions}
\end{figure}

Panel a) is a representaion of a possible "strip" of 4 slits.
Slit length is tuned so that the largest object is well
contained within the slit and a sky region on each side is allowed for.
In Figure 2b, we show how to accomodate spectra in the FOV. Each slit
(black thick line) produces a first order spectrum centred in Y around the slit, a zero order spectrum of few pixels right below, and
a second order spectrum above. 
The zero order consits in a very high, narrow (in the Y direction) peak.
Its contamination to underlying spectra cannot be removed, but, being not larger than few pixels, good reduction can be achieved by simply masking out
the affected pixels.
Contribution of
the second order is around only 10\%, but it is spread over
$\sim$ 600 pixels along the dispersion direction. Therefore contamination by second order must be handled carefully (see below). Making use of the fact that the FOV is 2048x2340 pixels, but the
CCD s are 2048x4096 pixels, we can place slits over the full 8 arcminutes in Y. In Figure 2 c,d,e,f we illustrate what are the contraints to be taken
into account when placing slits. In 2c, two slits of the same width
are placed, not aligned in Y direction. In this case the second order of the first slit (starting from the bottom)
partially covers the first order of the following slit. 
Such a situation makes background subtraction very difficult. In the second case
(2d), we have two slits aligned in Y, but the first being 
shorter than the second. Also in this case, good quality background subtraction from the second spectrum is impossible. In 2e we show
two slits of same length, aligned, but with the first order of the
first one partially superimposed to the first order of the second one.
It is obvious that data will be of bad quality.
Finally, in
2f, we put two slits of same length, aligned one to the other,
and positioned so that the second order of the first slit totally
covers the first order of the second slit. In this case
background subtraction is still possible with reasonable results.
To summarize, SPOC must take into account the following constraints: along
a vertical "strip" slits must be aligned and with same length. Their
distance in Y must be so that the first order spectra do not fall
one onto the other.

When designing SPOC, great attention must be paied to the possible
biases an automatization algorithm can introduce in observations. 
A first bias is intrinsic to
any MOS instrument: to allow for multiplexing, slits will have the tendency to align themselves in horizontal bands. But there are two more biases which can be easily introduced: if the algorithm starts its work
always from one point (e.g. top left) then more slits
will be placed there with respect to other parts of FOV, and, secondly, 
if the only criterium is to maximize the number of objects, and the
slit length is tuned on object size, smaller (i.e. fainter) objects
will be privileged.

Given all what above stated, we can now choose the algorithm. If we divide each
quadrant into subsequent vertical "strips" each strip being as large as the longest
possible slit, then this becomes a purely combinational problems.
Taking into account all possibilities sums up to
$~10^{73}$ different combinations to be computed to find the best one.
A simplification can be introduced by applying the same criteria as in the well known travelling salesman problem, i.e. consider only the most
probable solutions. For each X (spatial) position, we can vary
the strip width from a minimum (given by the space we allow for sky)
and a reasonable maximum (given by the largest object in the field we
are interested in). It is easy to show that the function 
(Number of slits)/(slit width) vs. the slit width has a maximum
for a paricular slit width. We can compute such a maximum for each
X position, and finally find the best combination of strip
widths which maximizes the total number of slits.

\section{Results}

\begin{figure}
\epsfxsize=9cm % fix the x-dimension and scales y-dim. to x-dim.
% Feel free to do the choice you prefer but do not exceed the x-dimension
% of the text lines
\hspace{1.5cm}\epsfbox{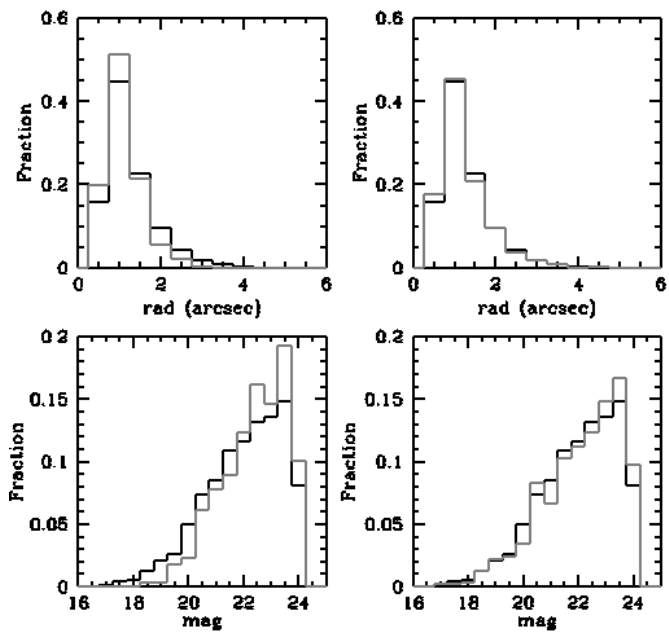} 
%for centering: act on hspace argument 
\caption[h]{VIMOS MPS results on simulated field}
\end{figure}

We have applied SPOC to both real data sets, and to artificial ones.
Computational time is negligible, being around 2 seconds on a
Sun Ultra Sparc 5 for one quadrant.
Apart from the intrinsic MOS bias, we have found that there is
still a tendency to privilege smaller objects with respect to bigger ones. To overcome this, we have introduced the possibility of
ignoring the object size to place slits in a strip, and
introducing it only when the maximum number of objects has been placed in a strip. The optimization in this case decreases by about 10\%
the maximum number of slits,
with the advantage that diameter and magnitude distributions
of the input catalogue and of the observed objects
are now statistically indistinguishable. Both possibilities (full optimization,
slightly biased and unbiased lower optimization) will be
offered to the user.

In Fig. 3 we show the results obtained by applying SPOC to a simultaed catalogue. The black line represents the input catalogue, and the grey line the
distributions obtained with SPOC (fully optimized in the left panels, unbiased
in the right panels). Panels a and b clearly show that the
revised version is unbiased against object size. The
difference in terms of number of objects is 5\% 
 
%\acknowledgements
%Type here the acknowledgements.

% References. We avoided using the \bibitem commmand since we found it is
% somewhat platform-dependent. We also avoided using the \cite{keyword}
% command since we found it cumbersome. However, if you are an expert 
% LateX user you may use the various LateX tools for the references 
% provided they give the same printout formats of the examples given here.

\end{document}